\documentclass[final,3p,times]{elsarticle}

\usepackage{amssymb}
\usepackage{amsmath}

\usepackage{hyperref}
\usepackage{cleveref}
\usepackage{xspace}
\usepackage{mathtools}
\usepackage{booktabs}
\usepackage{changepage}
\usepackage{multirow}

\usepackage{tikz}
\usetikzlibrary{fit,arrows}
\usetikzlibrary{calc}

\pgfdeclarelayer{aux}
\pgfdeclarelayer{hybrid}
\pgfdeclarelayer{mech}
\pgfdeclarelayer{pinn}
\pgfsetlayers{pinn,aux,hybrid,mech,main}

\newcommand{\dypopm}{\texttt{Dy\_PopMosq}\xspace}
\newcommand{\hypopm}{\texttt{Hy\_PopMosq}\xspace}

\crefname{equation}{Eq.}{Eqs.}

\journal{Ecological Modelling}

\begin{document}

\begin{frontmatter}



\title{Modelling Mosquito Population Dynamics using PINN-derived Empirical Parameters}


\author[1]{Branislava Lalic\corref{cor1}\fnref{fn1}}
\ead{branislava.lalic@polj.edu.rs}
\author[2]{Dinh Viet Cuong\fnref{fn1}}
\ead{dinhviet.cuong@dcu.ie}
\author[3]{Mina Petric}
\ead{mpetric@avia-gis.com}
\author[4]{Vladimir Pavlovic}
\ead{vladimir@pavlovic.net}
\author[1]{Ana Firanj Sremac}
\ead{ana.sremac@polj.edu.rs}
\author[5]{Mark Roantree}
\ead{mark.roantree@dcu.ie}

\cortext[cor1]{Corresponding author.}
\fntext[fn1]{Equal contribution.}

\affiliation[1]{organization={Faculty of Agriculture, University of Novi Sad}, city={Novi Sad}, country={Serbia}}
\affiliation[2]{organization={School of Computing, Dublin City University}, city={Dublin}, country={Ireland}}
\affiliation[3]{organization={Avia-GIS}, city={Zoersel}, country={Belgium}}
\affiliation[4]{organization={Department of Computer Science, Rutgers University}, city={New Brunswick, NJ}, country={USA}}
\affiliation[5]{organization={Insight Centre for Data Analytics, Dublin City University}, city={Dublin}, country={Ireland}}


\begin{abstract}
Vector-borne diseases continue to pose a significant health threat globally with more than 3 billion people at risk each year. Despite some limitations, mechanistic dynamic models are a popular approach to representing biological processes using ordinary differential equations where the parameters describe the different development and survival rates. Recent advances in population modelling have seen the combination of these mechanistic models with machine learning. One approach is physics-informed neural networks (PINNs) whereby the machine learning framework embeds physical, biological, or chemical laws into neural networks trained on observed or measured data. This enables forward simulations, predicting system behaviour from given parameters and inputs, and inverse modelling, improving parameterisation of existing parameters and estimating unknown or latent variables. In this paper, we focus on improving the parameterisation of biological processes in mechanistic models using PINNs to determine inverse parameters. In comparing mechanistic and PINN models, our experiments offer important insights into the strengths and weaknesses of both approaches but demonstrated that the PINN approach generally outperforms the dynamic model. For a deeper understanding of the performance of PINN models, a final validation was used to investigate how modifications to PINN architectures affect the performance of the framework.  By varying only a single component at a time and keeping all other factors constant, we are able to observe the effect of each change.
\end{abstract}




\end{frontmatter}



\section{Introduction}

Vector-borne diseases, transmitted by arthropods such as mosquitoes, pose a significant global health threat, with more than 500 million people infected annually \cite{WHO2023}. More than 3 billion people are at risk yearly, with mosquito-borne diseases comprising a substantial share of this burden. The transmission of VBDs is closely linked to vector population density, particularly in epidemic settings. Mosquitoes of the genera \emph{Aedes} and \emph{Culex} are vectors for multiple diseases, including West Nile fever, Saint Louis encephalitis, Japanese encephalitis, dengue and chikungunya \cite{becker2020mosquitoes}. Given their critical role in disease transmission, modelling the dynamics of the mosquito population is essential for predicting and controlling outbreaks. Mosquito population dynamics models have traditionally been classified into mechanistic (deterministic)~\cite{erickson2010a,cailly2012a,virgillito2021a,frantz2024a} and stochastic models~\cite{otero2006a,edwards2021a}. More recently, this classification has been expanded to include machine learning (ML) models (see, for example,~\cite{tsantalidou2021a,kinney2021a,joshi2021a,zhang2024a}).

Mechanistic dynamic models based on a cause-effect framework are widely used to model the dynamics of the mosquito population. These models typically represent biological processes using ordinary differential equations (ODEs), where the parameters describe the different development and survival rates. This approach offers strong interpretability, relatively low data requirements, and predictive power that extends beyond the data used for calibration and validation. However, this approach also presents several limitations, including the oversimplification of model parameters, often assuming linear relationships among variables, which can fail to accurately capture the complexities of biological systems, limited automatic adaptability to composite, real-time data (e.g., satellite imagery) and difficulties in capturing complex, multivariate interactions. Mechanistic models are mainly based on empirical parameters that are typically determined taking into account only the impact of meteorological conditions~\cite{tran2013a}, commonly temperature~\cite{erraguntla2021a}, and in some cases precipitation and relative humidity~\cite{yamana2013a}, to simulate the dynamics of the insect population. Factors such as land type and land cover characteristics, level of urbanisation availability of blood sources, vegetation dynamics, and availability of breeding sites, all of which significantly influence the mosquito population, are accounted for in only a few studies and typically only as a subset rather than in combination~\cite{dare2022a,yamashita2018a}.

Unlike mechanistic models, ML trained on extensive and high-quality datasets, including environmental, vegetation, and climatic variables, can achieve high precision in short-term predictions of mosquito population dynamics \cite{dare2025a,steindorf2025a}. However, these models typically require substantial computational resources, extensive training data, access to large-scale data storage, and technical expertise for model development and maintenance~\cite{joshi2021a}. Furthermore, ML models are inherently correlative and often lack explicit representation of biological processes, limiting their ability to simulate cause-effect relationships or to generalise beyond the conditions represented in their training data~\cite{athni2024a}. As a result, they can struggle to account for and interpret the impacts of different elements of climate change scenarios or environmental shifts without continual retraining or integration with mechanistic approaches~\cite{dare2025a,ferraguti2024a}.
A promising approach to advance insect population modelling is the combination of mechanistic and ML models. An example of this approach is physics-informed neural networks (PINNs)~\cite{karniadakis2021a}. They represent a novel ML framework that embeds physical, biological, or chemical laws into neural networks trained on observed and/or measured data. This allows both forward simulations, predicting system behaviour from given parameters and inputs, and inverse modelling, improving parameterisation of existing parameters and estimating unknown or latent variables~\cite{wesselkamp2024a}.

PINNs have recently gained popularity in biological sciences, where data is often scarce, noisy, or incomplete, but governing equations exist~\cite{oneto2025a,aatif2020a,lotfollahi2023a}. Furthermore, biological data are often combined with, for example, climate data, which are typically more accurate, introducing variable accuracy in the training data set, the so-called multifidelity data~\cite{karniadakis2021a,meng2019a}. Although mechanistic models struggle to account for varying data accuracy, PINNs are designed to handle these variations, making them advantageous for insect population modelling. By applying PINNs to this field, researchers can simultaneously predict populations and infer biologically significant data and knowledge, overcoming key limitations of traditional modelling methods~\cite{wesselkamp2024a}.

\textbf{Contribution.}
This paper is part of a larger study exploring the applicability of PINNs in the modelling of insect populations. We refer to our model as a physics-informed neural network; however, the underlying model is not strictly physics-based; instead, it is a mechanistic model rooted in biological principles, where ODEs effectively represent biology-informed constraints. Although \cite{cuong2024a} demonstrated the effectiveness of PINNs in obtaining forward solutions for mosquito populations, in this paper, we focus on improving the parameterisation of biological processes in the mechanistic dynamic model developed by~\cite{petric2020a}, for \emph{Culex} and \emph{Aedes species}, by using PINNs to determine inverse parameters. Throughout the rest of the paper, we will refer to it as the hybrid model.

%



The paper is structured as follows: 
\autoref{sec:methods} provides a brief overview of how PINNs work, introduces mechanistic dynamics, PINN, and hybrid models, describes the data, and explains the validation methodology;
\autoref{sec:results} presents \emph{Culex pipiens} population simulation results for Petrovaradin (Serbia) using mechanistic dynamic and hybrid models;
\autoref{sec:ablation} presents an ablation study in order to provide higher levels of interpretability for the results of experiments; 
\autoref{sec:discuss} discusses the advantages and limitations of our approach in comparison to similar studies; 
and finally, \autoref{sec:concl} presents potential avenues for future research.


\section{Methodology}
\label{sec:methods}

\subsection{PINNs for mosquito population dynamics: Forward and inverse modelling}\label{sec:pinn}

Although PINNs are increasingly used in biological sciences, their application, particularly for inverse problem solving in ecological modelling, remains relatively unfamiliar to many in the field. To address this, we briefly describe how PINNs work in the context of an ODE mosquito population dynamics model.

PINNs integrate entomological knowledge directly into their framework by embedding the governing equations of mosquito population dynamics into the neural network's loss function. This loss function consists of two primary components: the data loss term, measuring deviations between neural network predictions and observed mosquito population data; and the physics loss term, quantifying the discrepancies between neural network outputs and the governing ODEs. Despite its biological foundation, we retain the term \emph{physics loss} to maintain consistency with the standard PINN methodology. The data loss term ensures that the neural network predictions align closely with the observed data. The physics loss term ensures consistency with known~entomological dynamics, represented by the ODEs and the mosquito birth rates, mortality rates, and carrying capacities. The scaling coefficients for these two terms allow us to assign different weights to observations or ODEs, balancing the influence of data and physics constraints.

For forward problems, PINNs take initial conditions and \textbf{fixed} entomological parameters (such as birth rates, mortality rates, and carrying capacity) to train a neural network that predicts temporal changes in mosquito populations, guided by the physics loss and the data condition. Once trained to minimise the loss function, PINNs can~predict~mosquito population time series as long as the input data remains within the domain of the training input data.

For inverse problems, PINNs aim to infer unknown biological parameters from observed population data. The neural networks simultaneously learn the population dynamics and the parameters of the ODEs by minimising the data loss and physics loss functions. Once trained, the PINN can infer ODE parameters which can subsequently be used with traditional, mechanistic ODE models to simulate population dynamics.


Next, we introduce a hybrid approach aimed at refining the parameterisations of both the ground and empirical components within a mechanistic dynamic model. The model used in this study was constructed using the PINN model described in~\cite{cuong2024a}, allowing refined parameterisation through the application of the inverse theory framework.

\subsection{Mechanistic dynamic model (\dypopm)} 

We selected the ODE vector population dynamics model (\dypopm) developed in \cite{petric2020a} as a hybridisation case study. The mosquito life cycle is divided into ten stages: egg (\(E\)), larva (\(L\)), pupa (\(P\)), emerging adults (\(A_{\text{em}}\)), nulliparous blood-seeking adults (\(A_{b1}\)), nulliparous gestating adults (\(A_{g1}\)), nulliparous ovipositing adults (\(A_{o1}\)), parous blood-seeking adults (\(A_{b2}\)), parous gestating adults (\(A_{g2}\)) and parous ovipositing adults (\(A_{o2}\)). The dynamical system is expressed in \cref{eq:ode_sys} as a full ODE system where the details of all parameters are covered in~\autoref{tab:params} in \ref{sec:supa}. 

\begin{equation}
    \begin{cases}
    \frac{\text{dE}}{\text{dt}} = \gamma_{\text{Ao}}\left( \beta_{1}A_{o1} + \beta_{2}A_{o2} \right) - \left( \mu_{E} + f_{E} \right)E \\
    \frac{\text{dL}}{\text{dt}} = f_{E}E - \left( m_{L}\left( 1 + \frac{L}{\kappa_{L}} \right) + f_{L} \right)L \\
    \frac{\text{dP}}{\text{dt}} = f_{L}L - \left( m_{P} + f_{P} \right)P \\
    \frac{dA_{\text{em}}}{\text{dt}} = f_{P}\sigma e^{- \mu_{\text{em}}\left( 1 + \frac{1}{\kappa_{P}} \right)}P - \left( m_{A} + \gamma_{\text{Aem}} \right)A_{\text{em}} \\
    \frac{dA_{b1}}{\text{dt}} = \gamma_{\text{Aem}}A_{\text{em}} - \left( m_{A} + \mu_{r} + \gamma_{\text{Ab}} \right)A_{b1} \\
    \frac{dA_{g1}}{\text{dt}} = \gamma_{\text{Ab}}A_{b1} - \left( m_{A} + f_{\text{Ag}} \right)A_{g1} \\
    \frac{dA_{o1}}{\text{dt}} = f_{\text{Ag}}A_{g1} - \left( m_{A} + \mu_{r} + \gamma_{\text{Ao}} \right)A_{o1} \\
    \frac{dA_{b2}}{\text{dt}} = \gamma_{\text{Ao}}\left( A_{o1} + A_{o2} \right) - \left( m_{A} + \mu_{r} + \gamma_{\text{Ab}} \right)A_{b2} \\
    \frac{dA_{g2}}{\text{dt}} = \gamma_{\text{Ab}}A_{b2} - \left( m_{A} + f_{\text{Ag}} \right)A_{g2} \\
    \frac{dA_{o2}}{\text{dt}} = f_{\text{Ag}}A_{g2} - \left( m_{A} + \mu_{r} + \gamma_{\text{Ao}} \right)A_{o2}
    \end{cases}, \label{eq:ode_sys}
\end{equation}

These parameters are determined from observations and laboratory experiments and are considered a baseline for this research~\cite{petric2020a,focks1993a,erickson2010a,cailly2012a,ezanno2015a, Vinogradova1960}. The air temperature $\tau$ is considered the main forcing factor in \dypopm. The initial conditions are set to $300$ for all stages and the spin-up time of the model is estimated to be approximately $200$ days {[}26{]}. No bias reduction is applied in the original model.

We denote by \(\boldsymbol{u}(t) = \left( E(t),L(t),P(t),A_{\text{em}}(t),A_{b1}(t),A_{g1}(t),A_{o1}(t),A_{b2}(t),A_{g2}(t),A_{o2}(t) \right) \in \mathbb{R}^{10}\) the state of the mosquito dynamical system at time \(t\), where each element of \(u\) corresponds to the population count of a specific life stage in the mosquito life cycle. Let \(\boldsymbol{\theta}\) denote the system parameters defined in \cref{eq:ode_sys}. For brevity, we summarise the system as shown in \cref{eq:fode_system}.

\begin{equation}\label{eq:fode_system}
    \frac{d\boldsymbol{u}}{dt} = f_{ODE}(\boldsymbol{u},t; \boldsymbol{\theta}). 
\end{equation}

Going forward, we refer to individual ODEs in this system using \cref{eq:fode_j}, where $j=1,\ldots,10$.

\begin{equation}\label{eq:fode_j}
    \frac{du_j}{dt} = f_{ODE}^{(j)}(\boldsymbol{u},t; \boldsymbol{\theta}),
\end{equation}

\subsection{PINN Framework for Hybrid Mosquito Population Model (\hypopm)}
\label{subsubsec:pinn}
The PINN mosquito population dynamics model was developed based on the work presented in~\cite{cuong2024a}, which used only idealised annual daily temperature variations as input meteorological conditions. However, based on our knowledge of mosquito population development, we anticipated that daily air temperature and their extreme values, air humidity, and precipitation play an active role in the dynamics of the mosquito population. We introduce a novel aspect into \dypopm by employing a parameter network that models the variability of the \dypopm ODE parameters $\boldsymbol{\theta}$ as a function of a complete set of measured meteorological data. Next, we present the details of this new model, which we refer to as \hypopm.


\begin{equation}
\label{eq:thetann}
    \boldsymbol{\theta} \coloneq {\Theta}(\boldsymbol{m}; \boldsymbol{W}_\theta),
\end{equation}


Where \(\boldsymbol{m}(t)\) denote the vector of meteorological conditions at time \(t\), in \hypopm we assume that \cref{eq:thetann} models each unknown ODE system parameter \(\boldsymbol{\theta}\) by a neural network ${\Theta}$ with parameters $\boldsymbol{W}_\Theta$. The goal is to learn $\boldsymbol{W}_\Theta$ from historical observations $\boldsymbol{u}(t_i)$ paired with known meteorological conditions $\boldsymbol{m}(t_i)$.  We next present the framework for this by extending the approach proposed in~\cite{cuong2024a}.  

Assume we are given a set of observations \(D_{u} = \left\{ \left( t_{1},\boldsymbol{u}_{1} \right),\left( t_{2},\boldsymbol{u}_{2} \right),\ldots,\left( t_{j}, \boldsymbol{u}_{j} \right),\ldots \right\}\). The observations may be incomplete since only some elements of \(\boldsymbol{u}_j\) could be observed at each $t_i$. To train the neural network ${\Theta}$ we adopt a PINN approach, inspired by~\cite{cuong2024a,raissi2018a}. We introduce two auxiliary functions, the meteorological conditions auxiliary function $\boldsymbol{m}(t) \approx M(t; \boldsymbol{W}_M)$ and the state auxiliary function $\boldsymbol{u}(t) \approx \boldsymbol{u}_U(t) = U(t; \boldsymbol{W}_U)$ to facilitate, respectively, the generalisation of meteorological conditions $\boldsymbol{m}$ beyond the observed conditions, the need to avoid numerical integration to estimate the model state $\boldsymbol{u}$ during the learning process, and the differentiability necessary for the PINN learning framework.  Both auxiliary functions are modelled as neural networks, whose parameters are to be learnt during the PINN process described in \autoref{fig:pinn_framework_alt}. We assume that we have trained apriori \(M(t)\) that approximates the meteorological variables with reasonable precision.

\begin{figure}[!ht]
    \centering
    \resizebox{0.9\textwidth}{!}{\input{media/pinn_tikz_v2}}
    \caption{\textbf{Hybrid Mosquito Population Model}.
    \hypopm enhances \dypopm model by replacing its parameters with a neural network $\Theta$ learned through a PINN process. Two neural networks are trained jointly: state network $U(t;W_U)$ and the parameter network $\Theta(\boldsymbol{m};W_\Theta)$, where the driver $\boldsymbol{m} = M(t)$ are provided.
    The training minimizes a data loss on observations $\mathcal{D}_u = \left\{ (u_i, t_i) \right\}$ and a physics loss on collocation times $t_j$.
    Once trained, $\Theta$ yields parameters $\theta$ for forward simulation of the population $u(t)$.}
    \label{fig:pinn_framework_alt}
\end{figure}


\subsubsection{Neural Network Architecture} 
\begin{figure}[!ht]
    \centering
    \includegraphics[width=0.6\linewidth]{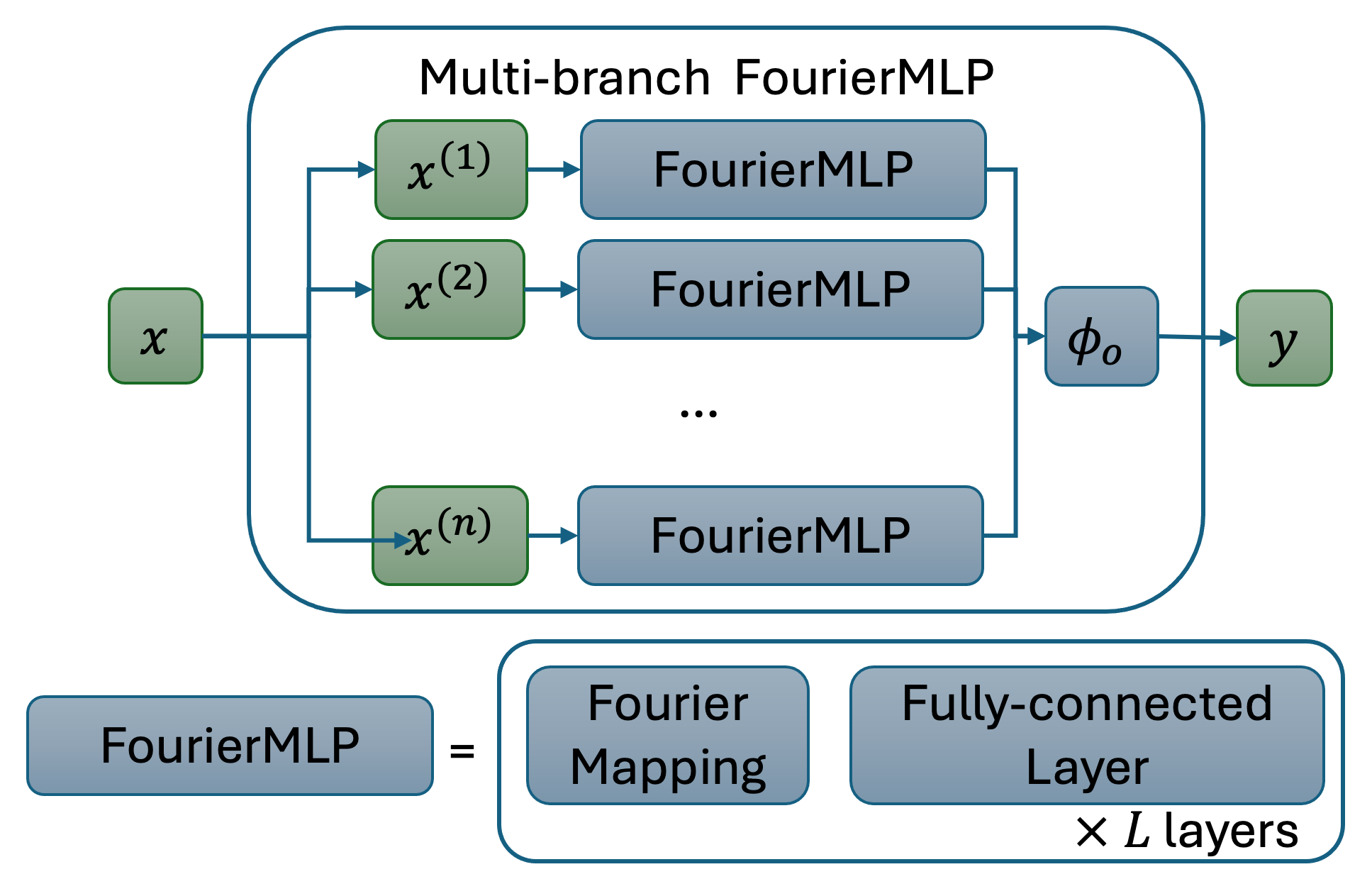}
    \caption{\textbf{Multi-branch Fourier-MLP PINN Framework.}
    Each branch receives a distinct group of inputs and passes them through a FourierMLP, which is a Fourier-feature layer followed by fully-connected layers. The branch outputs are summed and passed through the $SoftAbs$ activation function.}
    \label{fig:fouriermlp}
\end{figure}

For the neural network architectures of \(U,\ \Theta\) and possibly $M$, we implement FourierMLP \cite{tancik2020a} outlined in \autoref{fig:fouriermlp}, which has demonstrated significant improvements in both convergence speed and accuracy for PINN training~\cite{wang2021a}. Given an input \(x\) to the neural network FourierMLP, the output \(y\) is defined in ~\cref{eq:FourierMLP}, where \(B\) is a random matrix with entries sampled from a normal distribution \(N(0,\sigma)\).

\begin{align}
    h^{(0)} &= \left[ \cos \left( Bx \right) ; \sin \left( Bx \right) \right] \nonumber \\
    h^{(l)} &= \phi_h (W^{(l)} h^{(l-1)} + b^{(l)}), l=1,\dots,L-1   \label{eq:FourierMLP} \\
    y &= \phi_o \left( W^{(L)} h^{(L-1)} + b^{(L)} \right), \nonumber
\end{align}

 \(W^{(l)}\) and \(b^{(l)}\) represent the weights and biases of appropriate dimensions, respectively. During the training process, \(B\) remains fixed while the weights and biases are optimised. \(L\) denotes the number of hidden layers, and \(\phi_{h}\) is the element-wise activation function applied to the hidden layers. 

We employ the Gaussian Error Linear Unit (GeLU) activation function \cite{hendrycks2023a} for \(\phi_{h}\), as it provides a smooth non-linearity to the model.  To enforce non-negativity constraints on either state or parameters, we propose a soft absolute function as the output activation function \(\phi_{o}\) which we have called $SoftAbs$.
This novel function is shown in \cref{eq:softabs} where \(\epsilon\) is determined to be \(10^{- 4}\) during a rigorous validation process presented in \autoref{sec:act_fn}.

\begin{align}
    \phi (x) = \sqrt{x^2 + \epsilon} - \sqrt{\epsilon}, x \in \mathbb{R}
    \label{eq:softabs}
\end{align}

The meteorological variables \(A(t)\) typically originate from a high-dimensional space where the available training data is often limited. This frequently leads to poor generalisation of trained models. Therefore, we propose a multi-branch architecture for the external-to-parameter networks \(\Theta\), as illustrated in~\autoref{fig:fouriermlp}. Our proposed architecture comprises multiple branches (\(n\) branches are depicted in~\autoref{fig:fouriermlp}), each consisting of a separate FourierMLP that processes a distinct group of inputs. The outputs from these branches are subsequently combined in later layers to produce the final prediction.

Formally, consider \(n\) groups of external factors represented by vectors \(x^{(1)},\ldots,x^{(n)}\). Each input vector \(x^{(i)}\) is processed by a dedicated branch \(\text{FourierMLP}^{(i)}\). The outputs from these branches are then aggregated through summation and passed through a final activation function \(\phi_{o}\), as described in \cref{eq:MultiBranchFourierMLP}.

\begin{align}
    y = \phi_o \left( \sum_i \text{FourierMLP}^{(i)} \left(x^{(i)} \right) \right).
    \label{eq:MultiBranchFourierMLP}
\end{align}

The effectiveness of these novel architectures, the multi‑branch FourierMLP parameter network and the SoftAbs output activation, is systematically validated in \autoref{sec:ablation}.

\subsubsection{PINN-based Learning}
Neural networks \(U\) and \(\Theta\) are jointly optimised to minimise the objective function captured in~\cref{eq:loss_pinn,eq:loss_absfun,eq:loss_data}:
\begin{equation}\label{eq:loss_pinn}
L = L_{\text{data}} + L_{\text{ODE}} = L_{\text{data}} + \frac{1}{F}\sum_{i=1}^F \lambda_{i}L_{f^{(i)}}    
\end{equation}
with
\begin{align}
    L_{\text{data}} &= \frac{1}{N_{u}}\sum_{k=1}^{N_u} \lVert U\left( t_{k} \right) - \boldsymbol{u}_{k} \rVert^{2} \label{eq:loss_data}\\
    L_{f^{(i)}} &= \ \frac{1}{N_{f}} \sum_{j=1}^{N_f} \left( \left.\frac{dU^{(i)}}{dt}\right|_{t_j} - f^{(i)}_{ODE}\left( t_{j},U\left( t_{j} \right),\Theta\left( M\left( t_{j} \right) \right) \right) \right)^{2} \label{eq:loss_absfun}.
\end{align}
\(\lambda_{i}\) are weights that balance the multiple objectives within the loss function; \(N_{u}\) denotes the number of state observations; while \(N_{f}\) represents the number of collocation points \(t_{j}\) randomly sampled from the interval \(\lbrack 0,T\rbrack\). When the state is not fully observed, unavailable entries are masked out in the loss \(L_{\text{data}}\).

The objective function comprises two components: \(L_{\text{data}}\) that ensures that the predictions made by the model \(U\) closely align with the observed data, and \(L_{f^{(i)}}\) that minimises the residuals of the ODE to ensure that the model parametrised with the neural network aligns with the ODE system. By optimising these terms simultaneously, we aim to learn a system state and parameter set that not only fits the real-world observations but also satisfies the underlying ODE system.

We employ the gradient-based Adam optimiser~\cite{kingma2017a} to minimise the objective functions discussed earlier. All differentiation operations, including derivatives in the ODE equations and optimisation gradients, are computed using automatic differentiation provided by the PyTorch framework~\cite{paszke2019a}. To improve convergence and accuracy, we implement ODE normalisation and gradient balancing techniques as proposed in~\cite{cuong2024a}. ODE normalisation rescales the inputs and outputs of the neural networks and reformulates the loss function to ensure that these quantities remain within reasonable ranges. Gradient balancing adaptively adjusts the weights \(\lambda_{i}\) throughout the training process to maintain balance across tasks when optimising multiple objectives simultaneously. Furthermore, we resample the collocation points for ODE residual calculations at each training step from a uniform distribution over the time domain. To facilitate convergence, we initially train the networks using only the data loss, allowing the network \(U\) to capture the general solution shape before progressing to satisfying both objectives. Finally, we train a separate FourierMLP to approximate meteorological data, providing a continuous function \(M\) for the framework. This neural network is frozen during the PINN training process, serving as a fixed input to the main model.

\subsubsection{\hypopm specification}

Evaluation of model ground parameterisations (\ref{sec:supb}, \autoref{fig:sensitivity}) shows that the pupa development rate, \(f_{P}\) (or larva development rate \(f_{L} = 1.65f_{P}\)), is the key leverage point of the \dypopm model, exerting the greatest influence on the number of adults, both female and male (\(A_{b1} + A_{b2}\)). These results are consistent with entomological findings, which recognise pupa and larva development rates as critical factors influencing population growth dynamics~\cite{beck-johnson2013a, loetti2011a}. Therefore, we select \(f_{P}\) as the system parameter to be learned by the PINN model from meteorological measurements and entomological observations. Once learned, this parameter will replace empirical formulas in the \dypopm ODE system, creating the hybrid \hypopm model.

A custom-designed numerical experiment was created to address the key features of mosquito population dynamics: the timing and intensity of population (number of adults) growth, as well as the timing of population local peaks (temporary surge in population time series), which is, from the point of protective measures application, the most important population time series characteristic.

In addition to analysing the normalised model outputs, equal focus is given to its first derivative which indicates the rate of change. The timing of population peaks is also examined thoroughly. Based on the meteorological forcing data, the integration time step is set to 1 day. The initial conditions are predefined only for the first year using the same initial conditions learnt by the neural network \emph{U} in training data. For each subsequent year, the simulated values at the end of the previous year are used as initial conditions for the following year's simulations, ensuring the continuity condition of the population time series. Details of the numerical experiment design can be found in \ref{sec:appendix_exp_config}.

\section{Experiments}
\label{sec:results}
\subsection{Data}\label{sec:data}

Data used for training Hy\_PopMosq are obtained from a two-year experiment conducted at a semi-urban location in Petrovaradin (Serbia) during 2016 and 2017. Daily mosquito trap counts were recorded alongside meteorological measurements, including daily air temperature (average, maximum, minimum), relative humidity (average), and precipitation. The comparison study uses daily meteorological data to run both the \dypopm and \hypopm models for the period of 2000-2007. Model validation is based on daily mosquito trap counts at the same location, conducted once per week (referred to as weekly catches) during 2000-2007 period. As meteorological measurements were not conducted on-site during this time, data from 2016 and 2017 for Petrovaradin and the Rimski Sancevi (Serbia) climate station were used to create a linear regression model, which estimates values at the Petrovaradin location based on available Rimski Sancevi climate station data.

\subsection{Validation Methodology}\label{sec:valid}

Performance metrics used in the study are: a) root-mean-square-error (RMSE) and standard deviations of the simulated and observed (\(\sigma_{o}\)) normalised population, along with their first derivatives; and b) peak prediction metrics that include recall (percentage of observed peaks that are correctly predicted), precision (percentage of predicted peaks that match the observed peaks), and f1-score (harmonic average of~recall~and precision). The model is considered superior if, for normalised population values and their first derivatives, a) the RMSE is smaller and b) following \cite{Pielke1984}, the RMSE is less than the standard deviation of observed values (\(\sigma_{o}\)), and the standard deviation of the simulations closely matches the standard deviation of the observed values (\(\sigma_{o}\)). This combination of error minimisation (through RMSE) and variability matching (through \(\sigma\) comparison) helps to ensure both accuracy and reliability in the model's performance. When one model produces RMSE values smaller than \(\sigma_{o}\), while the \(\sigma\) of the simulated data is closer to \(\sigma_{o}\) for another model, we do not immediately conclude which model is better. Instead, we revisit both the observed and simulated data to investigate why simulations may have, for example, a smaller RMSE but a larger value for \(\sigma\).

\subsection{Results}

\begin{figure}[!ht]
    \centering
    \includegraphics[width=0.80\linewidth]{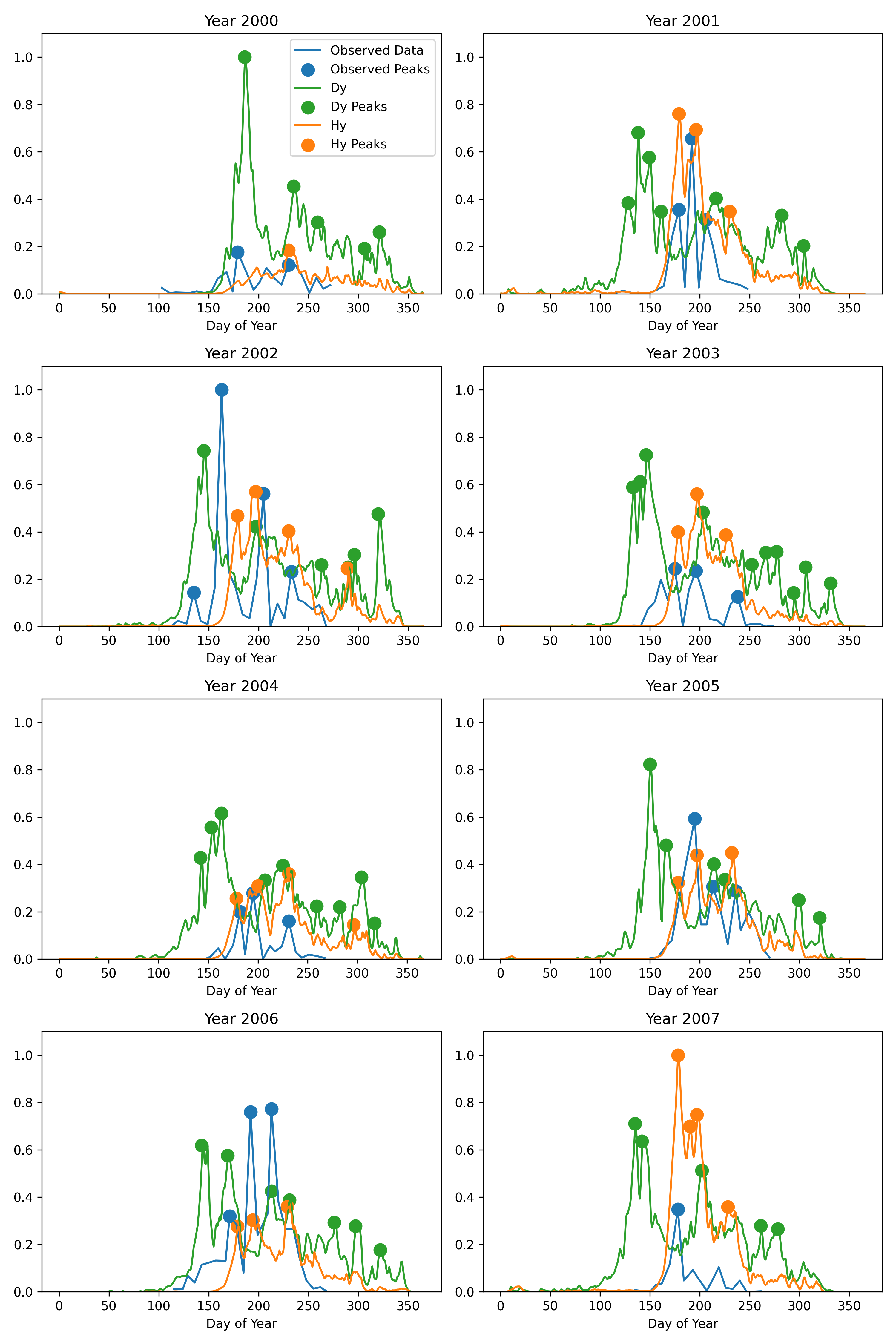}
    \caption{\textbf{Observed vs. simulated adult-mosquito abundance.}
    Daily, normalised counts of blood-seeking adults ($A_{b1}+A_{b2}$) are plotted for each calendar year.
    Circles mark peaks identified with a 7-day, 0.2-prominence detector.
    }
    \label{fig:valid_simulation_ab}
\end{figure}

\begin{figure}[!ht]
    \centering
    \includegraphics[width=0.80\linewidth]{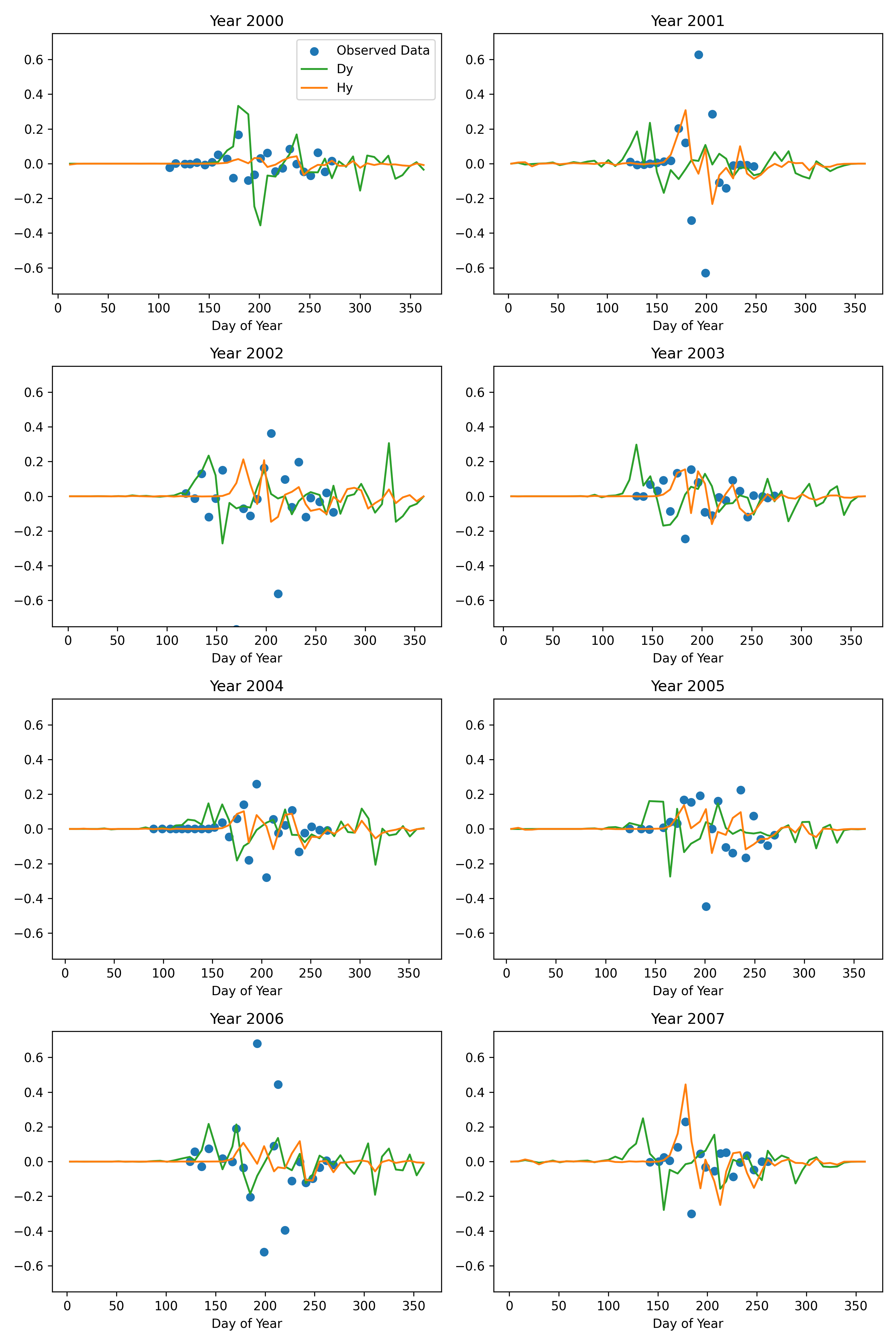}
    \caption{\textbf{Rate of adult mosquito population growth.}
    Each panel shows the first difference (week-to-week increment) of the normalised adult population timeseries depicted in \autoref{fig:valid_simulation_ab}.}
    \label{fig:valid_simulation_diff}
\end{figure}

The comparison of \hypopm and \dypopm models, based on the performance metrics (\autoref{tab:tab1} and \autoref{tab:annual_performance}) and the simulated and observed values of the adult mosquito population (\autoref{fig:valid_simulation_ab}), as well as the rate of adult mosquito population growth (\autoref{fig:valid_simulation_diff}), offers important insights into model strengths and weaknesses in simulating mosquito population dynamics. Both models are evaluated using outlined performance metrics and criteria (\ref{sec:valid}), focusing on error minimisation and variability matching. This analysis highlights the distinctive behaviour of models over the integration period that includes eight years, paying special attention to 2002 and 2007, where both models exhibited notable deviation from observed values. One of the key findings is that \hypopm generally outperforms \dypopm in terms of population RMSE, achieving an overall lower RMSE (0.18 on average compared to 0.26 for \dypopm). This is not lower than \(\sigma_o = 0.13\), but is closer to this threshold in comparison to \dypopm. The lower RMSE values are found across all years (\autoref{tab:annual_performance}), including the outlier years 2002 and 2007, when the observed population peaks contribute to high RMSE values for both models. The peak in June 2002 (\autoref{fig:valid_simulation_ab}) exceeds typical counts of adult mosquitoes, suggesting that environmental factors outside of meteorological conditions, such as pond retention after heavy rain or a water pipe burst, significantly affected population dynamics. Despite these deviations, \hypopm performs better in terms of capturing population trends and maintaining a more accurate alignment with observed data.

\begin{table}[!ht]
\centering
\caption{\textbf{Performance metrics for \dypopm and \hypopm.} Values in brackets refer to performance metrics when two worst-performing years are removed.}
\label{tab:tab1}
\begin{tabular}{llll}
\hline
 & \dypopm & \hypopm & \textbf{Observed} \\
\hline
RMSE\_ population & 0.26 (0.25) & 0.18 (0.14) & - \\
$\sigma\_ population$ & 0.16 & 0.12 & 0.13 \\
RMSE\_ rate & 0.21 (0.17) & 0.17 (0.14) & - \\
$\sigma\_ rate$ & 0.12 & 0.10 & 0.16 \\
No. Peaks & 3.63 & 1.38 & 1.88 \\
Peak recall & 0.13 & 0.56 & \\
Precision Peak & 0.09 & 0.63 & \\
F1 Peak & 0.11 & 0.57 & \\
\hline
\end{tabular}
\end{table}
When comparing both models' ability to match observed population variability, \hypopm proves to be closer to the standard deviation of observed values (\(\sigma\) = 0.12) than \dypopm (\(\sigma\) = 0.16), further supporting its superiority in matching population fluctuations.

In simulating population growth rates (\autoref{fig:valid_simulation_diff}), \hypopm once again demonstrates a stronger overall performance with an average growth rate RMSE of 0.17 compared to 0.21 for \dypopm and standard deviation of observed population growth rate of 0.16 (\autoref{tab:tab1} and \autoref{tab:annual_performance}). This indicates that \hypopm is more accurate in simulating the rate of change in mosquito populations. However, both models have difficulties in replicating the sharp population changes registered in 2007, indicating that limitations remain in both models' capacity to capture rapid shifts in population dynamics. Additionally, when analysing the growth rate variability, \dypopm exhibits a slight advantage (\(\sigma\_ rate\)= 0.12) over \hypopm (\(\sigma\_ rate\)= 0.10), but not enough to affect our determination regarding model superiority.

An additional insight emerges when excluding the two worst-performing years, 2002 and 2007, from the analysis. When removed, \hypopm's performance shows a more substantial improvement over \dypopm's performance, as demonstrated by the values in brackets in \autoref{tab:tab1}. This indicates \hypopm's robustness and ability to handle typical population dynamics better than \dypopm, which struggles with both extreme and normal conditions.

Peak detection is another element in which \hypopm outperforms \dypopm. The \hypopm model achieves a higher peak recall (0.56) and precision (0.63), leading to an F1 score of 0.57, which is significantly better than \dypopm's performance, where peak recall (0.13) and precision (0.09) are much lower. \autoref{fig:valid_simulation_ab} visually supports the findings, showing that \hypopm more closely aligns with the peaks in the observed adult population compared to \dypopm.

This indicates that enhanced parameterisation of other development rates can lead to more accurate simulations of mosquito population dynamics and peak predictions. It is important to note that beyond the commonly used average daily air temperature, the PINN model used a full set of meteorological data during the learning process. Furthermore, both models demonstrated numerical stability. While determining initial conditions based on the end-of-previous year's results is physically realistic, it carries the risk of error propagation in long-term simulations.






\section{Model Architecture Analysis}
\label{sec:ablation}
Using the same experimental procedure as in \autoref{sec:methods}, we now investigate how the proposed addition of neural network architectures and the activation function affect the performance of the framework. 
Specifically, we first train PINNs using data from the training period, during which the model learns how meteorological variables affect the mosquito parameters.
Then, the trained parameter network is used to predict parameters for the test period, and these parameters are substituted into the ODE system to simulate the mosquito population representing the population predictions.
The difference is that we vary one component at a time while keeping all other factors constant to observe the effect of each change.
As a form of ablation study, we first examine the architecture of the neural networks and then, analyse the non-negative activation function.

\subsection{Model Architectures}

To understand the contribution of each neural network component, we conduct an ablation study on two architectural choices: the use of Fourier feature layers and the branched structure.
We consider four combinations of these two elements:
\begin{enumerate}
    \item \textbf{MLP}: All of the neural networks for the system state $U$ and system parameters $\Theta$ use standard MLPs, similar to those in conventional PINNs~\cite{raissi2018a}.
    \item \textbf{FourierMLP}: All of the neural networks $U$ and $\Theta$ use FourierMLPs, similar to those in~\cite{wang2021a}.
    \item \textbf{Branched MLP}: For comparison, we use a branched version of the MLP, which is similar to the multi-branch Fourier MLP described in \autoref{subsubsec:pinn}, but the Fourier layer is replaced with a normal fully connected layer, making each branch a standard MLP.
    This multi-branch MLP is used for the parameter networks $\Theta$, while the state network remains an MLP. 
    \item \textbf{Branched FourierMLP}: Finally, our proposed architecture, which is the same as in \autoref{subsubsec:pinn}, uses a FourierMLP for the network $U$ and a multi-branch FourierMLP for the parameters $\Theta$.
\end{enumerate}

As PINN accuracy does not depend heavily on the capacities of the neural networks~\cite{krishnapriyan2021characterizing}, it was only necessary to experiment with a single hyperparameter setting for each configuration.
Hyperparameters  for all neural networks are set as follows: for all the state networks, we use a network with four hidden layers; the first hidden layer has 256 units, and the three subsequent hidden layers each have 128 units.
For the parameter networks, the non-branched versions use a similar setting: one layer of 256 units and three layers of 128 units. 
For the branched versions, each branch uses half the number of units, that is, 128 units for the first hidden layer and 64 units for the other three.
All other hyperparameters are identical to those in \ref{sec:appendix_exp_config}, including the use of the GELU activation function for hidden layers and the soft absolute function with $\epsilon = 10^{-4}$ for the output.

\autoref{tab:pinn2:model_architecture_errors} presents the simulation error metrics obtained when simulating the mosquito dynamical models using the parameter $f_P$ learned from the four different neural network configurations.
It can be seen that PINNs across all architectures outperform the empirical formula.
The Branched FourierMLP used in \autoref{sec:methods} achieves the best overall performance, exhibiting the lowest RMSE and the highest scores in peak detection metrics.
When comparing the standard MLP to its branched counterpart, we observe that the RMSE values are similar ($0.1937$ for MLP vs $0.1979$ for Branched MLP), but the Branched MLP exhibits marginal improvements in peak detection metrics. 
Specifically, the peak recall  increases from $0.1458$ to $0.2708$, the precision peak from $0.3750$ to $0.4375$, and the peak F1-score from $0.2083$ to $0.2917$.
This suggests that the branching architecture enhances the model's capacity to capture features relevant to peak occurrences, such as annual patterns.
Interestingly, the standard MLP outperforms the FourierMLP in terms of RMSE ($0.1937$ vs $0.2479$), suggesting that the FourierMLP may be overfitting the data due to its higher complexity.
However, when the FourierMLP is integrated into a branched architecture, as in Branched FourierMLP, the model not only mitigates overfitting but also leverages the Fourier features to capture periodic patterns more effectively.
The Branched FourierMLP achieves a lower RMSE of $0.1791$ compared to both the standard MLP and the FourierMLP, and significantly improves peak detection metrics over both models.
These results suggest that the branching architecture allows the neural network to generalize better while the Fourier features enable it to capture periodic components in the data. 
The combination between the branching structure and Fourier features in the Branched FourierMLP contributes to its superior performance, making it a robust framework for learning system parameters in PINNs.

\begin{table}[!ht]
\centering
\caption{\textbf{Error Metrics from Parameters learned from PINNs comprising different network architectures}. The best results for each metric are highlighted in \textbf{bold}, while the second best results are \underline{underlined}.}
\label{tab:pinn2:model_architecture_errors}
{%
\begin{tabular}{lrrrr}
\hline
                     & MLP      & FourierMLP & Branched MLP & Branched FourierMLP \\ \hline
RMSE                 & \underline{0.193658} & 0.247858   & 0.197923         & \textbf{0.179100}       \\
Diff RMSE     & \underline{0.187189} & 0.207561   & 0.191858         & \textbf{0.172800}       \\
2nd Diff RMSE & 0.314825 & 0.348473   & \underline{0.312093}         & \textbf{0.292700}                \\
Recall Peak          & 0.145833 & 0.166667   & \underline{0.270833}         & \textbf{0.562500}                \\
Precision Peak       & 0.375000 & 0.085714   & \underline{0.437500}         & \textbf{0.625000}                \\
F1 Peak              & 0.208333 & 0.101010   & \underline{0.291667}         & \textbf{0.566700}                \\ \hline
\end{tabular}}
\end{table}

\subsection{Activation Functions for Non-negativity}
\label{sec:act_fn}

\begin{figure}[!ht]
    \centering
    \includegraphics[width=0.6\linewidth]{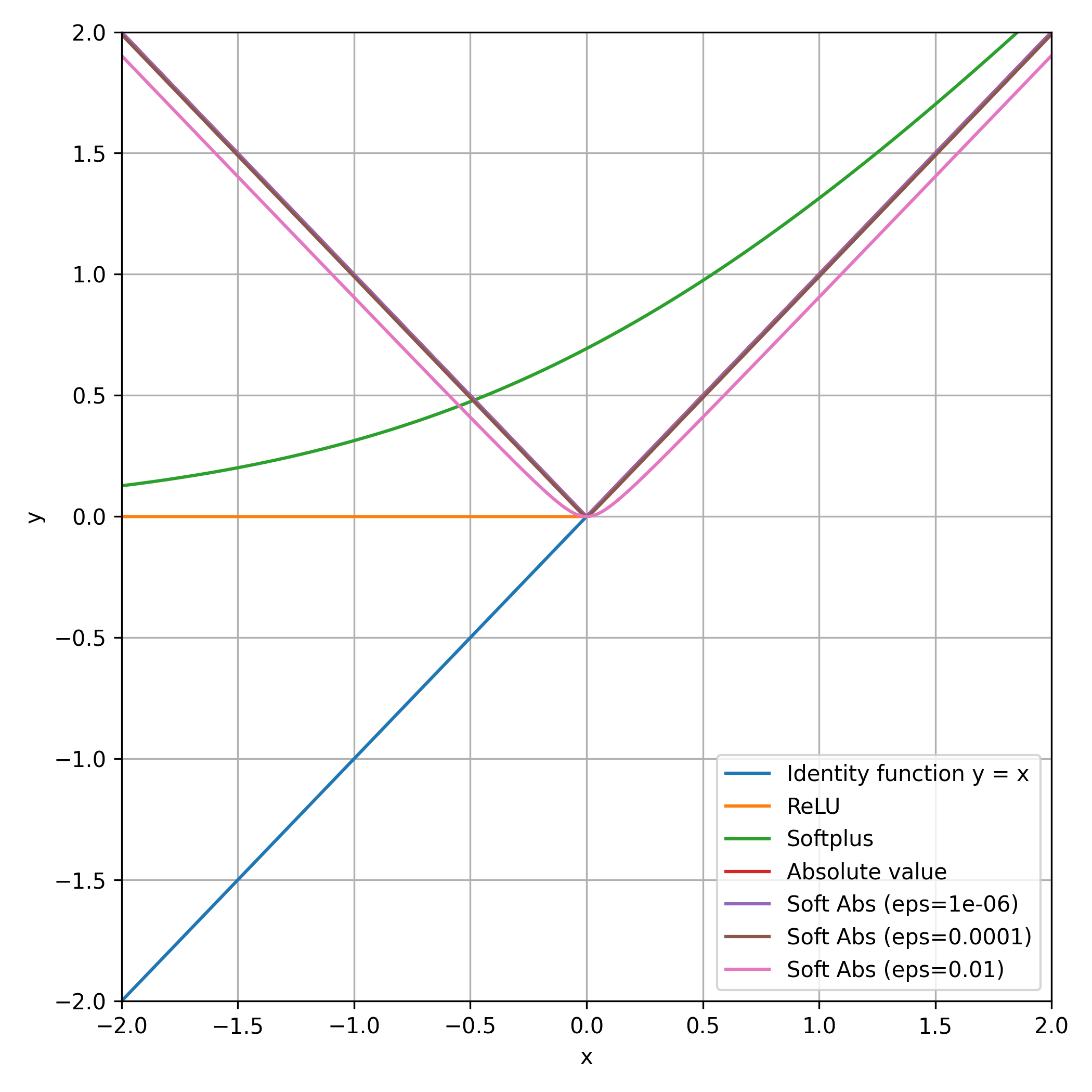}
    \caption{\textbf{Non-negativity Activation Functions}. The Soft Abs functions with  $\epsilon=10^{-6}$ and $\epsilon=10^{-4}$ appear very close in the plot, closely resembling the positive part of ReLU and the identity function.}
    \label{fig:pinn2:activation_function}
\end{figure}

In the second part of this evaluation, the aim is to explore different activation functions to enforce the non-negativity of system parameters and states. 
In \autoref{fig:pinn2:activation_function}, each function is plotted.
The baseline activation function is the identity function, which returns the exact input with which it was provided.
Other popular activation functions that express the non-negativity property are ReLU and its smoothed version, Softplus~\cite{glorot2011deep}.
While ReLU and Softplus are popular activation functions, they suffer from the critical drawback of causing dying neurons~\cite{lu2019dying}.
This phenomenon occurs when the input to the ReLU is negative, resulting in zero gradient propagation and causing the neuron to stop learning.
This problem is exacerbated when applied to PINNs.
In \cite{cuong2024a}, it was confirmed that PINNs may converge to the (trivial) zero solution, which is precisely how dying neurons are triggered.
This convergence causes PINNs to become stuck at a local minimum and prevents further learning.

To address this issue, we propose using the absolute value function to impose non-negativity on the states and parameters.
The absolute value function effectively fixes the zero-gradient problem by ensuring a non-zero gradient in the negative domain.
However, the original absolute value function makes it difficult for the model to optimize values to zero, as the gradient is discontinuous at 0.0.
Therefore, we propose using a soft version of the absolute value function, which we refer to as  $SoftAbs$, defined in \cref{eq:softabs}.
When $\epsilon$ is small, the differences from the standard absolute value function are minimal, as seen in \autoref{fig:pinn2:activation_function}, but $SoftAbs$ is significantly more effective, as will be demonstrated in the evaluation.

\autoref{tab:pinn2:model_actfunc_errors} shows the error metrics obtained from simulations using parameters learned by PINNs with different output activation functions. 
The first 4 columns present the error metrics for the Identity, ReLU, Softplus and Abs functions while the remaining 3 columns present the metrics for $SoftAbs$ using 3 different parameter settings for $\epsilon$.
Overall, the $SoftAbs$ activation function with $\epsilon=10^{-4}$ demonstrates superior performance across multiple metrics.
It achieves the highest scores in all three peak detection metrics, and ranks second in RMSE and third in first-order difference error.
Other configurations of the $SoftAbs$ function also exhibit significant improvements, particularly in peak detection and second-order difference errors.
For example, $SoftAbs$ with $\epsilon=10^{-6}$ attains the lowest second-order difference RMSE of 0.2853 and achieves the second-highest peak F1-score of 0.5083.
Abs improves peak detection compared to ReLU and the identity function; however, its performance in RMSE metrics is limited.
These results suggest that softening the derivatives of the absolute value function aids the model in better detecting peaks while maintaining comparable RMSE values.
ReLU achieves the lowest first-order difference RMSE at 0.1706, while the identity activation function comes second in overall RMSE with a value of 0.1637.
This performance can be attributed to our re-selection procedure, which involves a simulating training period after initial training.

\begin{table}[!ht]
\centering
\caption{\textbf{Effect of output activation functions on the performance of PINN-learned parameters}. The best results for each metric are highlighted in bold face, while the second best results are underlined.}
\label{tab:pinn2:model_actfunc_errors}
{%
\begin{tabular}{lrrrrrrr}
\toprule
 & \multicolumn{7}{c}{\textbf{Activation function}} \\ \cmidrule(lr){2-8}
 & Identity & ReLU & Softplus & Abs & \multicolumn{3}{c}{SoftAbs ($\varepsilon$)} \\ \cmidrule(lr){6-8}
 & & & & & $10^{-6}$ & $10^{-4}$ & $10^{-2}$ \\ \midrule
RMSE           & 
\underline{0.1637}   & 0.1868 & 0.1981   & 0.2008 & 0.1903 & 0.1791 & \textbf{0.1558} \\
Diff RMSE      & 0.1768   & \textbf{0.1706} & 0.1916   & 0.1787 & 0.1748 & \underline{0.1728} & 0.1809 \\
2nd Diff RMSE  & 0.3031   & \underline{0.2927} & 0.3138   & 0.3080 & \textbf{0.2853} & \underline{0.2927} & 0.3194 \\
Recall Peak    & 0.2708   & 0.0416 & 0.2083   & 0.2500 & \underline{0.4583} & \textbf{0.5625} & 0.3333 \\
Precision Peak & 0.2708   & 0.1250 & 0.3125   & \underline{0.4375} & \textbf{0.6250} & \textbf{0.6250} & 0.5000 \\
F1 Peak        & 0.2500   & 0.0625 & 0.2083   & 0.3005 & \underline{0.5083} & \textbf{0.5667} & 0.3625 \\ \hline
\end{tabular}}
\end{table}

\begin{figure}[!ht]
    \centering
    \includegraphics[width=1.0\linewidth]{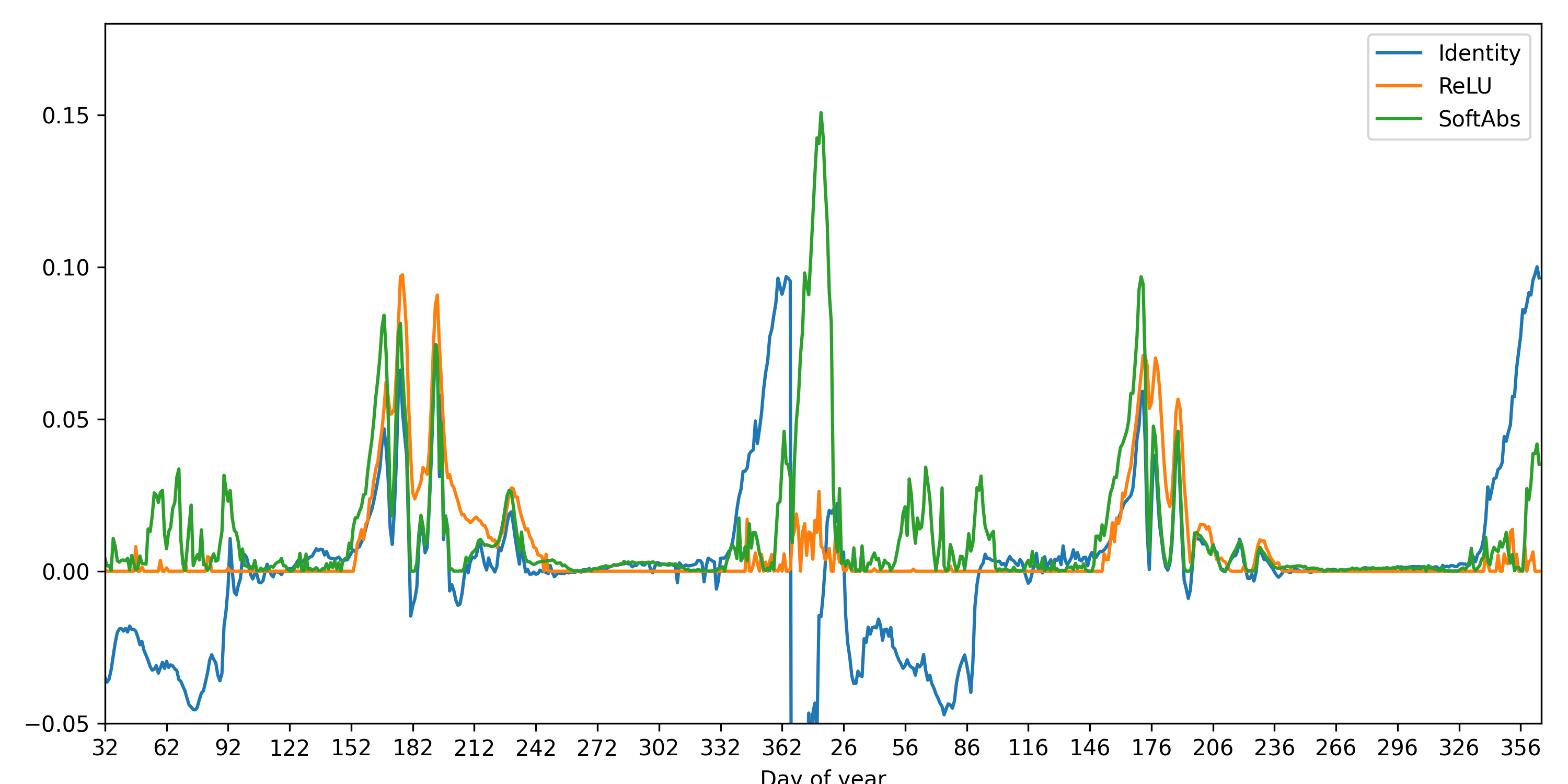}
    \caption{Parameter $f_P$ predictions with different output activation functions, training period}
    \label{fig:pinn2:act_func_fP}
\end{figure}

For a comparison of model predictions, \autoref{fig:pinn2:act_func_fP} plots the predictions of the learned pupa development rate $f_P$ over the training period using three different activation functions: Identity, ReLU, and SoftAbs with $\epsilon=10^{-4}$.
In the intervals from day 160 to 210 in the first year and from day 140 to 200 in the second year, all three models agree approximately on the values learned.
This convergence is attributed to the high mosquito activity during these periods, which results in increased observational mosquito counts and thus, provides information for the models to learn.
The critical differences among the models lie in the approximate range of days 50 to 90 in both years.
During these periods, the $SoftAbs$ activation function effectively captures the small peaks in mosquito counts by predicting surges in the corresponding $f_P$.
In contrast, the ReLU activation function encounters difficulties in learning values near zero, with predictions remaining mostly zero during these intervals.
This issue arises due to the dying neuron phenomenon, where neurons become inactive and are unable to learn once their outputs reach zero.
The Identity activation function fails to satisfy the non-negative constraints in the physical modeling of mosquito populations, making its predictions unrealistic.
Nevertheless, the Identity function achieves the lowest RMSE on the training data, a performance attributable to its relaxed constraints, which allow for a better fit to the data despite violating physical plausibility.

It is important to note that the models are selected based on the simulated population by \dypopm on the training data with PINN-trained parameters.
The chosen checkpoint may not be the fully converged checkpoints of the training process.
Models using ReLU and Softplus activation functions are chosen at 11,000 and 12,000 out of a total of 300,000 training steps, when the training loss had not yet been fully minimised.
In the long term, these models usually converge to the null solution, which means the output becomes zero.
In this case, the model satisfies the physical constraints very well, but it does not match the observed data.
This observation suggests that the optimised models for these two activation functions may reside at local minima in the loss landscape.
In contrast, models employing the Identity, Abs, and SoftAbs activation functions were selected at later stages of training, around 250,000 out of 300,000 training steps, where the training loss was minimised more thoroughly.

\section{Discussion}\label{sec:discuss}

The results obtained in this study highlight the advantages and limitations of both traditional mechanistic dynamical models (\dypopm) and their hybrid counterpart enhanced with physics-informed neural networks (\hypopm). Mechanistic dynamic models have been widely used to simulate mosquito population dynamics due to strong interpretability and reliance on biological principles. Studies such as ~\cite{erickson2010a} and ~\cite{focks1993a}, have successfully applied stage-structured population models to predict mosquito abundance and potential outbreak risks under varying climatic conditions. However, as noted by~\cite{cailly2012a, ezanno2015a}, traditional mechanistic models often fail to capture the full complexity of real-world mosquito population because they rely on simplified parameterisations of biological processes and environmental interactions. These models mainly incorporate temperature as a key forcing variable, neglecting other meteorological and ecological variables that could significantly impact mosquito development and survival. This limitation is evident in the results presented in Sec. 3, where \dypopm struggled to replicate population peaks and exhibited higher RMSE values in some years.

The integration of PINN-learned parameters in \hypopm improved the simulation accuracy by dynamically adjusting development rates based on a broader set of meteorological data, including extreme air temperature, relative air humidity and precipitation. This aligns with findings by~\cite{beck-johnson2013a, loetti2011a}, which emphasise the role of moisture availability in influencing larval habitat suitability and adult emergence rates. The performance metrics indicate that \hypopm consistently outperforms \dypopm in terms of: a) lower RMSE values for adult population and growth rate; b) improved peak prediction with higher recall and precision; and c) better alignment with observed seasonal trends, particularly during typical meteorological conditions. By integrating learned relationships from observed and measured data, \hypopm bridges the gap between existing parameterisations and empirical knowledge, effectively applying the learning-from-data concept.

While \hypopm demonstrates clear advantages, certain limitations remain.

\begin{enumerate} \def\labelenumi{\roman{enumi})}
\item \begin{quote} Neither model accurately predicts population peaks in 2002 and 2007, likely due to unaccounted environmental changes such as temporary water retention or infrastructure changes. While permanent changes can be introduced in an advanced training dataset, sudden, short-term environmental changes caused by human activities cannot and they will continue to be the source of model error. These limitations align with challenges identified in~\cite{ezanno2015a}, where models struggled to simulate sudden population fluctuations. \end{quote}
\item \begin{quote} Initial sensitivity analysis (\ref{sec:supb}, \autoref{fig:sensitivity}) shows that the pupa and larva development rates significantly impact adult population dynamics. Future improvements of \hypopm could refine determination of other parameters and enhance obtained results using additional environmental and human activity data. \end{quote}
\item \begin{quote} The Hybrid approach requires increased computational power for training and optimisation, which may limit their accessibility in resource-constrained settings.
This study prioritises framework development and accuracy and thus, we intentionally over-set hyperparameters such as network size, the number of training steps, the number of collocation points, etc. to ensure robust training.
Future work may explore more efficient configurations or pruning strategies to lower computational demands without compromising accuracy.
\end{quote} \end{enumerate}

Recent studies demonstrate the potential of combining dynamic models with neural networks beyond PINNs. Both forward and inverse simulations can be further improved by directly incorporating specific biological constraints into the neural network framework, as is commonly done in physics-guided neural networks~\cite{jia2021a}. This is particularly important when, for example, threshold values of meteorological elements determine the dynamics of specific biological processes. Additionally, it is important to consider that the training dataset may include variables that, while not used for parameterisation of biological processes in ODEs, are known to affect population dynamics. Such variables can significantly enhance forward and inverse simulations, particularly in parameter determination.

One example ~\cite{zhang2024a} is where integrated deep learning is used together with a mechanistic model to determine the relationship between oviposition rate of the \emph{Aedes} mosquito, with temperature and precipitation. Their approach combines deep learning-based parameter inference with a mechanistic ODE model, similar to \hypopm, but using a pure data-driven framework rather than physics-informed constraints. Similarly,~\cite{nisar2024a} proposed a recurrent neural network-based model coupled with ODE to predict mosquito-born disease outbreak.

\section{Conclusions}\label{sec:concl}
The goal of this work was to investigate the efficacy of PINN models to improve the parameterisation of biological processes and specifically, to determine inverse parameters.  The results of this study demonstrate that hybrid mechanistic-ML models offer a robust framework for improving insect population dynamics modeling. By integrating the interpretability of the ODE-based approach with the adaptability of data-driven model, \hypopm successfully addresses key weaknesses of traditional models, enhancing both predictive accuracy and generalisability. The ability of \hypopm to predict population peaks with higher accuracy is particularly relevant for vector control programs. Studies such as \cite{focks1993a} and \cite{petric2020a} emphasise the importance of early warning systems for targeted interventions. Improved forecasting through PINN-enhanced mechanistic models could allow for timely deployment of treatments or community-based interventions.

In addition, the ablation study provided insights into the performance of different PINN architectures, whereby varying only a single component at a time while keeping all other factors constant, we could observe the effect of each change. In particular, this study highlighted how the combination between the branching structure and Fourier features in the Branched FourierMLP model facilitates its superior performance, making it a robust framework for learning system parameters in PINNs and our choice for the methodology presented in \autoref{sec:methods}.

Further research will explore: the integration of remote sensing data (satellite data for example) to take the environmental impact of mosquito population dynamics into account; human population dynamics; and, the adaptation of \hypopm for other mosquito species, particularly Aedes vectors responsible for Dengue and Zika transmission. Beyond mosquito population modelling, the applicability of hybrid models makes them highly valuable in agricultural entomology, particularly in integrated pest management. By incorporating both mechanistic insights with observed and measured data, these models can improve pest outbreak predictions, optimise intervention strategies, and contribute to more sustainable and precise control practices.

\section{Data availability}

The datasets generated during and/or analysed during the current study are available in the GitHub repository at \href{https://github.com/dinhvietcuong1996/external-pinn-mosquito}{https://github.com/dinhvietcuong1996/external-pinn-mosquito}.

\section{Code Availability}

The code used for generating and analysing the data during the current study is available from the corresponding author upon reasonable request.


 \bibliographystyle{elsarticle-num} 
 \bibliography{main}

\begin{thebibliography}{10}
\expandafter\ifx\csname url\endcsname\relax
  \def\url#1{\texttt{#1}}\fi
\expandafter\ifx\csname urlprefix\endcsname\relax\def\urlprefix{URL }\fi
\expandafter\ifx\csname href\endcsname\relax
  \def\href#1#2{#2} \def\path#1{#1}\fi

\bibitem{WHO2023}
O.~H. et. al., \href{doi: 10.3390/insects14030221.}{Mosquito-borne diseases and their control strategies: An overview focused on green synthesized plant-based metallic nanoparticles.}, Insects 14(3) (2023) PMID: 36975906.
\newline\urlprefix\url{doi: 10.3390/insects14030221.}

\bibitem{becker2020mosquitoes}
N.~Becker, D.~Petri{\'c}, M.~Zgomba, C.~Boase, M.~B. Madon, C.~Dahl, A.~Kaiser, Mosquitoes: identification, ecology and control, Springer Nature, 2020.

\bibitem{erickson2010a}
R.~Erickson, S.~Presley, L.~Allen, K.~Long, S.~Cox, A stage-structured, aedes albopictus population model, Ecological Modelling 221  1273–1282.

\bibitem{cailly2012a}
P.~Cailly, A.~Tran, T.~Balenghien, G.~L’Ambert, C.~Toty, P.~Ezanno, A climate-driven abundance model to assess mosquito control strategies, Ecological Modelling 227  7–17.

\bibitem{virgillito2021a}
C.~Virgillito, M.~Manica, G.~Marini, B.~Caputo, A.~Della~Torre, R.~Rosà, Modelling arthropod active dispersal using partial differential equations: the case of the mosquito aedes albopictus, Ecological Modelling 456  109658.

\bibitem{frantz2024a}
R.~Frantz, H.~Godinez, K.~Martinez, W.~Cuello, C.~Manore, Age structured partial differential equations model for culex mosquito abundance, Ecological Modelling 494  110764.

\bibitem{otero2006a}
M.~Otero, H.~Solari, N.~Schweigmann, A stochastic population dynamics model for aedes aegypti: Formulation and application to a city with temperate climate, Bulletin of Mathematical Biology 68  1945–1974.

\bibitem{edwards2021a}
C.~Edwards, E.~Crone, Estimating abundance and phenology from transect count data with glms, Oikos 130  1335–1345.

\bibitem{tsantalidou2021a}
A.~Tsantalidou, E.~Parselia, G.~Arvanitakis, K.~Kyratzi, S.~Gewehr, A.~Vakali, C.~Kontoes, Mamoth: An earth observational data-driven model for mosquitoes abundance prediction, Remote Sensing 13  2557.

\bibitem{kinney2021a}
A.~Kinney, S.~Current, J.~Lega, Aedes-ai: Neural network models of mosquito abundance, PLoS Computational Biology 17  1009467.

\bibitem{joshi2021a}
A.~Joshi, C.~Miller, Review of machine learning techniques for mosquito control in urban environments, Ecological Informatics 61  101241.

\bibitem{zhang2024a}
M.~Zhang, X.~Wang, S.~Tang, Integrating dynamic models and neural networks to discover the mechanism of meteorological factors on aedes population, PLoS Computational Biology 20  1012499.

\bibitem{tran2013a}
A.~Tran, G.~L’Ambert, G.~Lacour, R.~Benoît, M.~Demarchi, M.~Cros, P.~Cailly, M.~Aubry-Kientz, T.~Balenghien, P.~Ezanno, A rainfall- and temperature-driven abundance model for aedes albopictus populations, International Journal of Environmental Research and Public Health 10  1698–1719.

\bibitem{erraguntla2021a}
M.~Erraguntla, D.~Dave, J.~Zapletal, K.~Myles, Z.~Adelman, T.~Pohlenz, M.~Lawley, Predictive model for microclimatic temperature and its use in mosquito population modeling, Scientific Reports 11.

\bibitem{yamana2013a}
T.~Yamana, E.~B. Eltahir, Incorporating the effects of humidity in a mechanistic model of anopheles gambiae mosquito population dynamics in the sahel region of africa, Parasites \& Vectors 6.

\bibitem{dare2022a}
D.~Da~Re, W.~Bortel, F.~Reuss, R.~Müller, S.~Boyer, F.~Montarsi, S.~Ciocchetta, D.~Arnoldi, G.~Marini, A.~Rizzoli, G.~L’Ambert, G.~Lacour, C.~Koenraadt, S.~Vanwambeke, M.~Marcantonio, Dynamaedes: a unified modelling framework for invasive aedes mosquitoes, Parasites \& Vectors 15.

\bibitem{yamashita2018a}
W.~Yamashita, S.~Das, G.~Chapiro, Numerical modeling of mosquito population dynamics of aedes aegypti, Parasites \& Vectors 11.

\bibitem{dare2025a}
D.~Da~Re, G.~Marini, C.~Bonannella, F.~Laurini, M.~Manica, N.~Anicic, A.~Albieri, P.~Angelini, D.~Arnoldi, F.~Bertola, B.~Caputo, C.~Liberato, A.~Della~Torre, E.~Flacio, A.~Franceschini, F.~Gradoni, P.~Kadriaj, V.~Lencioni, I.~Del~Lesto, F.~Russa, R.~Lia, F.~Montarsi, D.~Otranto, G.~L’Ambert, A.~Rizzoli, P.~Rombolà, F.~Romiti, G.~Stancher, A.~Torina, E.~Velo, C.~Virgillito, F.~Zandonai, R.~Rosà, Modelling the seasonal dynamics of aedes albopictus populations using a spatio-temporal stacked machine learning model, Scientific Reports 15.

\bibitem{steindorf2025a}
V.~Steindorf, H.~B, N.~Stollenwerk, A.~Cevidanes, J.~Barandika, P.~Vazquez, A.~García-Pérez, M.~Aguiar, Forecasting invasive mosquito abundance in the basque country, spain using machine learning techniques, Parasites \& Vectors 18.

\bibitem{athni2024a}
T.~Athni, M.~Childs, C.~Glidden, E.~Mordecai, Temperature dependence of mosquitoes: Comparing mechanistic and machine learning approaches, PLoS Neglected Tropical Diseases 18  0012488.

\bibitem{ferraguti2024a}
M.~Ferraguti, S.~Argany, C.~Mora-Rubio, D.~Bravo-Barriga, F.~Lope, A.~Marzal, Landscape and climatic factors shaping mosquito abundance and species composition in southern spain: A machine learning approach to the study of vector ecology, Ecological Informatics 102860.

\bibitem{karniadakis2021a}
G.~Karniadakis, I.~Kevrekidis, L.~Lu, P.~Perdikaris, S.~Wang, L.~Yang, Physics-informed machine learning, Nature Reviews Physics 3  422–440.

\bibitem{wesselkamp2024a}
M.~Wesselkamp, N.~Moser, M.~Kalweit, J.~Boedecker, C.~Dormann, Process‐informed neural networks: A hybrid modelling approach to improve predictive performance and inference of neural networks in ecology and beyond, Ecology Letters 27.

\bibitem{oneto2025a}
L.~Oneto, D.~Chicco, Eight quick tips for biologically and medically informed machine learning, PLoS Computational Biology 21  1012711.

\bibitem{aatif2020a}
M.~Aatif, J.~Tiwari, A facile approach for enhancing device performance of excitonic solar cells with an innovative sno2/tcne electron transport layer, AIP Advances 10.

\bibitem{lotfollahi2023a}
M.~Lotfollahi, S.~Rybakov, K.~Hrovatin, S.~Hediyeh-Zadeh, C.~Talavera-López, A.~Misharin, F.~Theis, Biologically informed deep learning to query gene programs in single-cell atlases, Nature Cell Biology.

\bibitem{meng2019a}
X.~Meng, G.~Karniadakis, A composite neural network that learns from multi-fidelity data: Application to function approximation and inverse pde problems, Journal of Computational Physics 401  109020.

\bibitem{cuong2024a}
D.~Viet~Cuong, B.~Lalić, M.~Petrić, N.~Thanh~Binh, M.~Roantree, \href{https://doi.org/10.1371/journal.pone.0315762}{Adapting physics-informed neural networks to improve ode optimization in mosquito population dynamics}, PLOS ONE 19~(12) (2024) 1--30.
\newblock \href {https://doi.org/10.1371/journal.pone.0315762} {\path{doi:10.1371/journal.pone.0315762}}.
\newline\urlprefix\url{https://doi.org/10.1371/journal.pone.0315762}

\bibitem{petric2020a}
M.~Petric, Modelling the influence of meteorological conditions on mosquito vector population dynamics (diptera, culicidae.

\bibitem{focks1993a}
D.~Focks, D.~Haile, E.~Daniels, G.~Mount, Dynamic life table model for aedes aegypti (diptera: Culicidae): Analysis of the literature and model development, Journal of Medical Entomology 30  1003–1017.

\bibitem{ezanno2015a}
P.~Ezanno, M.~Aubry-Kientz, S.~Arnoux, P.~Cailly, G.~L’Ambert, C.~Toty, T.~Balenghien, A.~Tran, A generic weather-driven model to predict mosquito population dynamics applied to species of anopheles, culex and aedes genera of southern france, Preventive Veterinary Medicine 120  39–50.

\bibitem{Vinogradova1960}
E.~B. Vinogradova, Experimental investigation of the ecological factors causing diapause of the adults of blood-sucking mosquitoes (diptera, culicldae), Entomol. Obozr. 39 (1960) 327--340.

\bibitem{raissi2018a}
M.~Raissi, P.~Perdikaris, G.~Karniadakis, Physics-informed neural networks: A deep learning framework for solving forward and inverse problems involving nonlinear partial differential equations, Journal of Computational Physics 378  686–707.

\bibitem{tancik2020a}
M.~Tancik, P.~Srinivasan, B.~Mildenhall, S.~Fridovich-Keil, N.~Raghavan, U.~Singhal, R.~Ramamoorthi, J.~Barron, R.~Ng, Fourier features let networks learn high frequency functions in low dimensional domains, Neural Information Processing Systems 33  7537–7547.

\bibitem{wang2021a}
S.~Wang, H.~Wang, P.~Perdikaris, On the eigenvector bias of fourier feature networks: From regression to solving multi-scale pdes with physics-informed neural networks, Computer Methods in Applied Mechanics and Engineering 384  113938.

\bibitem{hendrycks2023a}
D.~Hendrycks, K.~Gimpel, \href{http://arxiv.org/abs/1606.08415}{Gaussian error linear units (gelus}, preprint at.
\newline\urlprefix\url{http://arxiv.org/abs/1606.08415}

\bibitem{kingma2017a}
D.~Kingma, J.~Ba, \href{http://arxiv.org/abs/1412.6980}{A method for stochastic optimisation}, preprint at.
\newline\urlprefix\url{http://arxiv.org/abs/1412.6980}

\bibitem{paszke2019a}
A.~Paszke, S.~Gross, F.~Massa, A.~Lerer, J.~Bradbury, G.~Chanan, T.~Killeen, Z.~Lin, N.~Gimelshein, L.~Antiga, A.~Desmaison, A.~Kopf, E.~Yang, Z.~DeVito, M.~Raison, A.~Tejani, S.~Chilamkurthy, B.~Steiner, L.~Fang, J.~Bai, C.~S, \href{http://arxiv.org/abs/1912.01703}{Pytorch: An imperative style}, preprint at.
\newline\urlprefix\url{http://arxiv.org/abs/1912.01703}

\bibitem{beck-johnson2013a}
L.~Beck-Johnson, W.~Nelson, K.~Paaijmans, A.~Read, M.~Thomas, O.~Bjørnstad, The effect of temperature on anopheles mosquito population dynamics and the potential for malaria transmission, PLoS ONE 8  79276.

\bibitem{loetti2011a}
V.~Loetti, N.~Schweigmann, N.~Burroni, Development rates, larval survivorship and wing length ofculex pipiens(diptera: Culicidae) at constant temperatures, Journal of Natural History 45  2203–2213.

\bibitem{Pielke1984}
P.~Pielke, Mesoscale meteorological modeling, Academic Press, New York, N. Y.

\bibitem{krishnapriyan2021characterizing}
A.~S. Krishnapriyan, A.~Gholami, S.~Zhe, R.~M. Kirby, M.~W. Mahoney, Characterizing possible failure modes in physics-informed neural networks (2021).
\newblock \href {http://arxiv.org/abs/2109.01050} {\path{arXiv:2109.01050}}.

\bibitem{glorot2011deep}
X.~Glorot, A.~Bordes, Y.~Bengio, Deep sparse rectifier neural networks, in: Proceedings of the fourteenth international conference on artificial intelligence and statistics, JMLR Workshop and Conference Proceedings, 2011, pp. 315--323.

\bibitem{lu2019dying}
L.~Lu, Y.~Shin, Y.~Su, G.~E. Karniadakis, Dying relu and initialization: Theory and numerical examples, arXiv preprint arXiv:1903.06733 (2019).

\bibitem{jia2021a}
X.~Jia, J.~Willard, A.~Karpatne, J.~Read, J.~Zwart, M.~Steinbach, V.~Kumar, Physics-guided machine learning for scientific discovery: An application in simulating lake temperature profiles, ACM/IMS Transactions on Data Science 2  1–26.

\bibitem{nisar2024a}
K.~Nisar, M.~Anjum, M.~Raja, M.~Shoaib, Design of a novel intelligent computing framework for predictive solutions of malaria propagation model, PLoS ONE 19  0298451.

\bibitem{petri2020a}
D.~Petrić, R.~Bellini, E.-J. Scholte, L.~Rakotoarivony, F.~Schaffner, Monitoring population and insecticide resistance of aedes albopictus in europe, Parasite. Vector 13  1–20.

\end{thebibliography}

\section{Acknowledgement}

This research is partially supported by a grant from Research Ireland (Grant No. SFI/12/RC/2289\_P2) and by the European Union (Grant Agreement No.
101136578). The views and opinions expressed in this publication are
solely those of the author(s) and do not necessarily represent the
official position of the European Union or the European Commission.
Neither the European Union nor the granting authority is responsible for
any use that may be made of the information contained herein. Additional
support was provided by the Ministry of Science, Technological Development and Innovation of the Republic of Serbia through two Grant Agreements with the University of Novi Sad, Faculty of Agriculture (No. 451-03-66/2024-03/200117, dated February 5, 2024). 

The authors express their gratitude to Dušan Petrić, Aleksandra
Ignjatović Ćupina, Mihaela Kavran, Nikola Nožinić and Dragan Dondur for
their invaluable efforts in collecting the \emph{Culex pipiens} data
used in this study. The data were collected with financial support from
the City of Novi Sad, and the VectorNet project (ECDC/EFSA). We would
also like to acknowledge the work done within the WNED-X project of the
ISIDORe JRA programme, funded by the European Union's Horizon Europe
research and innovation programme under grant agreement number
101046133.

\section{Author contributions}

BL- design of dynamical systems hybridisation concept and main paper
writing; DVC-contributed to solution design and development, performed
all PINN numerical experiments, contributed to main paper writing; BL
and DVC equally contributed to the realisation of this research;
MP-provided meteorological and biological data and mosquito population
model; MR -~ contributed to solution design and overall supervision; VP
-- provided guidance on the conceptualisation and application of PINNs
in dynamic models hybridisation; all authors reviewed the paper.



\listoffigures

\listoftables

\appendix

\section{ODE Model Parameters}\label{sec:supa}
Detailed in \autoref{tab:params}

\begin{table}[!ht]
    \centering
    \caption{ODE Model parameters}
    \label{tab:params}
\begin{adjustwidth}{0cm}{}
\begin{tabular}{llll}
\toprule
Parameter & Description & Value & Unit \\
\midrule
\(\tau\) & Temperature & & \({^\circ}C\) \\
\(\gamma_{\text{Aem}}\) & Development rate of emerging adults & 1.143 &
\(\text{day}s^{- 1}\) \\
\(\gamma_{\text{Ab}}\) & Development rate of bloodseeking adults & 0.885
& \(\text{day}s^{- 1}\) \\
\(\gamma_{\text{Ao}}\) & Ovipositing adult development rate & 2 &
\(\text{day}s^{- 1}\) \\
\(f_{E}( > 0)\) & Egg development rate &
\(0.16 \cdot \left( e^{\left\lbrack 0.105(\tau - 10) \right\rbrack} - e^{\left\lbrack 0.105(38 - 10) - \frac{1}{5.007}(38 - \tau) \right\rbrack} \right)\)
& \(\text{day}s^{- 1}\) \\
\(f_{P}\) & Pupa development rate &
\(0.021 \cdot \left( e^{\left\lbrack 0.162(\tau - 10) \right\rbrack} - e^{\left\lbrack 0.162(38 - 10) - \frac{1}{5.007}(38 - \tau) \right\rbrack} \right)\)
& \(\text{day}s^{- 1}\) \\
\(f_{L}\) & Larva development rate & \(f_{P} \cdot 1.65\) &
\(\text{day}s^{- 1}\) \\
\(f_{\text{Ag}}( > 0)\) & Development rate of gestating adults &
\(\frac{\tau - 9.8}{64.4}\) & \(\text{day}s^{- 1}\) \\
\(m_{E}\) & Egg mortality rate & \(m_{E} = \mu_{E}\) &
\(\text{day}s^{- 1}\) \\
\(m_{L}\) & Larva mortality rate &
\(\exp\exp\ \left\lbrack - \frac{\tau}{2} \right\rbrack + \mu_{L}\ \) &
\(\text{day}s^{- 1}\) \\
\(m_{P}\) & Pupa mortality rate &
\(\exp\exp\ \left\lbrack - \frac{\tau}{2} \right\rbrack + \mu_{P}\ \) &
\(\text{day}s^{- 1}\) \\
\(m_{A}\ \left( > \mu_{A} \right)\) & Mortality rate for \(A_{b}\) &
\(- 0.005941 + 0.002965 \cdot \tau\) & \(\text{day}s^{- 1}\) \\
\(\mu_{E}\) & Minimum egg mortality rate & 0 & \(\text{day}s^{- 1}\) \\
\(\mu_{L}\) & Minimum larva mortality rate & 0.0304 &
\(\text{day}s^{- 1}\) \\
\(\mu_{P}\) & Minimum pupa mortality rate & 0.0146 &
\(\text{day}s^{- 1}\) \\
\(\mu_{\text{em}}\) & Mortality rate during emergence & 0.1 &
\(\text{day}s^{- 1}\) \\
\(\mu_{r}\) & Mortality rate during bloodseeking & 0.08 &
\(\text{day}s^{- 1}\) \\
\(\mu_{A}\) & Minimum adult mortality rate & 1/43 &
\(\text{day}s^{- 1}\) \\
\(\kappa_{L}\) & Carrying capacity for larva & \(8 \cdot 10^{8}\) &
\(\text{day}s^{- 1}\) \\
\(\kappa_{P}\) & Carrying capacity for pupa & \(10^{7}\) &
\(\text{day}s^{- 1}\) \\
\(\sigma\) & Sex ratio at emergence & 0.5 & - \\
\(\beta\) & Number of eggs per \(A_{o}\) &
\(\beta_{1} = 141\ \left( np^{*} \right),\ \beta_{2} = 80\ \left( p^{*} \right)\)
& - \\
\bottomrule
\end{tabular}
*np = nulliparous, p = parous
\end{adjustwidth}
\end{table}

\section{Parameter Sensitivity}\label{sec:supb}

To assess the most influential parameters in the mosquito dynamical system (\cref{eq:ode_sys}), we conduct a sensitivity analysis by
systematically varying parameters and observing their impact on the
system state. We examine 10 parameters:
\(\gamma_{\text{Aem}},\gamma_{\text{Ab}},\gamma_{\text{Ao}},f_{E},f_{P},f_{L},f_{A}g,m_{L},m_{P}\),
and \(m_{A}\), all of which are listed in \autoref{tab:params}.
Each parameter is individually altered at all time points \(t\) by \(- 10\%,\  - 5\%,\  + 5\%\), and \(+ 10\%\) of its original value, while keeping all other parameters unchanged. The Python ODE Solver is employed to simulate the system under these new conditions. The root mean squared error (RMSE) is then calculated between the new \(A_{b1} + A_{b2}\) values and those obtained without parameter changes.

\begin{figure}[!ht]
    \centering
    \includegraphics[width=0.98\linewidth]{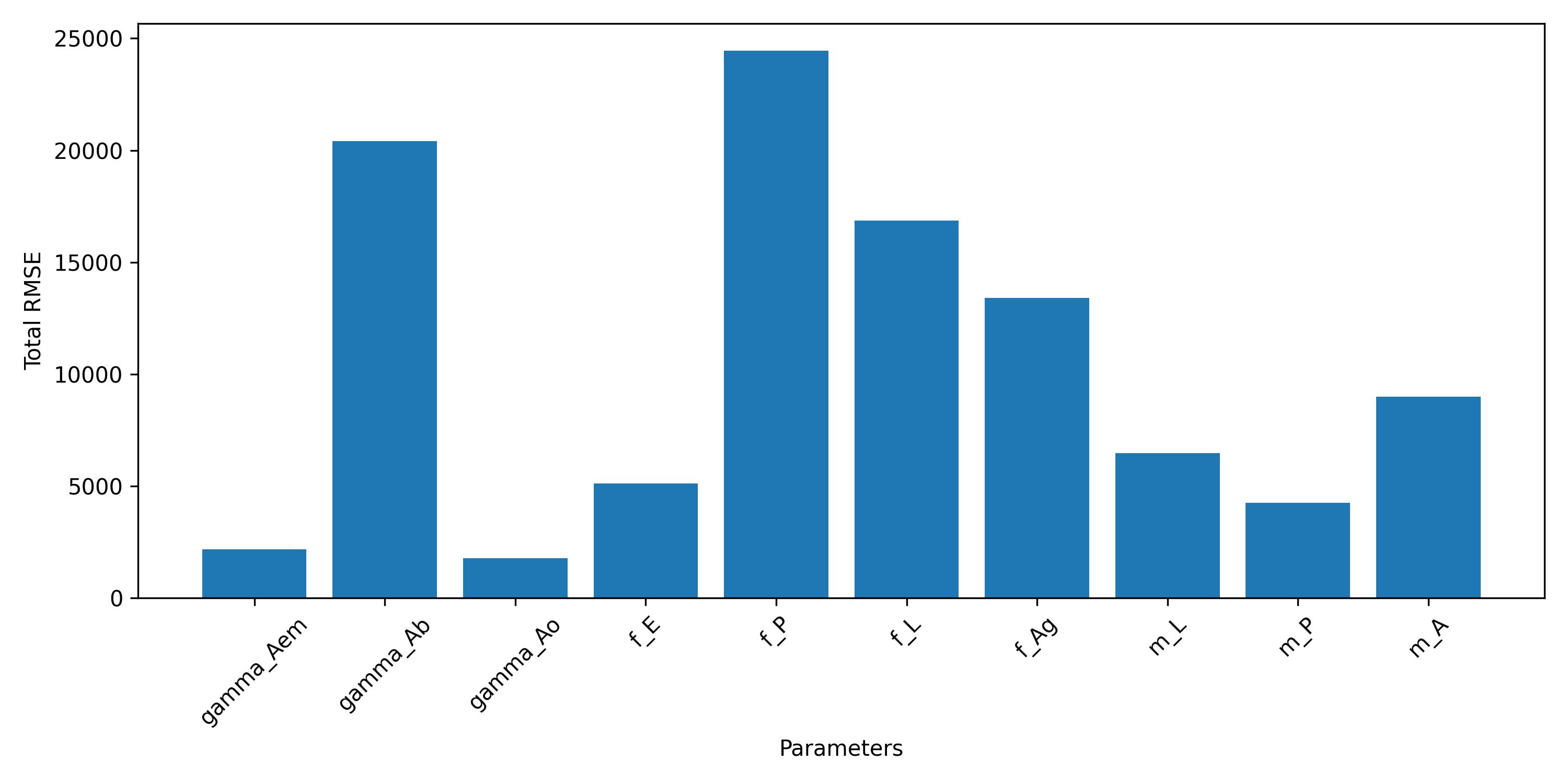}
    \caption{Parameter Sensitivity.}
    \label{fig:sensitivity}
\end{figure}

\autoref{fig:sensitivity} illustrates the average RMSE across the four alteration
levels for each parameter. The results indicate that the pupa
development rate \(f_{P}\) exerts the most significant influence on the
adult blood-seeking mosquito population \(A_{b1} + A_{b2}\). The
development rate of blood-seeking adults \(f_{\text{Ab}}\), which
directly affects \(A_{b1}\) and \(A_{b2}\) in the equations,
demonstrates slightly less impact. Parameters \(f_{\text{Ag}}\) and
\(m_{A}\) show marginal effects, while other parameters such as
\(\gamma_{\text{Aem}}\) and \(\gamma_{\text{Ao}}\) exhibit minimal
influence on the \(A_{b1} + A_{b2}\) quantity.

\section{Experiment Configuration}
\label{sec:appendix_exp_config}

The experiment begins with the pre-processing of mosquito data and the
determination of lower and upper bounds for all data columns, essential
for ODE Normalisation. We conduct a simulation using parameters derived
from empirical formulas based on climate condition data. The resulting
state values from this simulation are used to establish the bounds for
the system state. We rescale the collected mosquito counts to align with
this range and apply a 5-day window Spline smoothing to mitigate noise
in the data. Furthermore, the parameter values obtained from empirical
formulas determine their respective ranges while the bounds for climate
data are established through their available measurements.

For training using this 2-year dataset, the function \(M\), defined over
the time domain, plays a critical role. We train a FourierMLP to interpolate meteorological measurements for any real-valued time \(t\)
within the domain. This model is configured with 256 Fourier features,
followed by three hidden layers, each comprising 128 units. The model
outputs two groups of external features: one group consisting of three
meteorological variables (temperature, humidity, and precipitation), and
another representing the day of the year, with values ranging from 0 to
365.

The neural network architectures are designed as follows: The system
state neural network \(U\) comprises 256 Fourier features followed by
three hidden layers of 128 units each. For the parameter network, we
implement a two-branch FourierMLP. The first branch considers 7-day
historical meteorological data as input, while the second branch takes
day-of-week as input.
This configuration allows the second branch to capture inherent annual
patterns of the system, while the former learns the effects of
meteorological conditions on mosquito development rates. Both networks
employ GELU activation functions for hidden layers and the soft absolute
function (\cref{eq:softabs}) to enforce non-negativity.

We train neural network \(M\) for 300,000 epochs, saving the checkpoint
with the lowest root mean squared error (RMSE). For PINN training, we
initially train with only data loss for 10,000 steps, followed by
290,000 steps with the full objective function. PINN checkpoints are
saved every 500 steps, and we select the checkpoint yielding the best
RMSE when simulating with PINN-learned parameters. The selected
checkpoint serves as the final model and is validated using the
validation dataset.

We present our results by comparing simulations using PINN-learned
parameters against those using baseline parameters, via graphical
representations and quantitative metrics. For the PINN-learned
simulations, we use the same initial conditions derived from the trained
network \(U\) for both the training period and the 7-year validation
dataset. The baseline simulations use an initial condition of 300 for
each state vector component, consistent with the work in \cite{petri2020a}.
To facilitate comparison, all simulation results and observations are
normalised to the range \(\lbrack 0,1\rbrack\) for the calculation of
metrics. Our validation metrics include the root mean squared error
(RMSE) between \(A_{b1} + A_{b2}\), the RMSE of the weekly difference in
\(A_{b1} + A_{b2}\), and the 7-day 0.2-prominence peak detection recall,
precision, and F1-score.

The experiment is implemented in Python, with neural networks and
optimisations written in PyTorch. Training is accelerated using a
GeForce GTX 4090 GPU. All simulations are executed using SciPy {[}S1{]},
using its finite-difference ODE solver {[}S2{]}. The peak detection
algorithm is also part of the same package.

\section{Annual performance metrics for \dypopm and \hypopm}

\begin{table}[!ht]
  \caption{Annual performance metrics for \dypopm\ and \hypopm}
  \label{tab:annual_performance}
  \centering
  \begin{tabular}{llrrrrrr}
    \toprule
    {Model} & {Year} &
    {RMSE} & {$\sigma$} & {$\sigma_o$} &
    {RMSE\_rate} & {$\sigma_{\text{rate}}$} & {$\sigma_{o,\text{rate}}$} \\
    \midrule
    \multirow{9}{*}{\textbf{\hypopm}}
        & 2001 & 0.21 & 0.17 & 0.16 & 0.25 & 0.16 & 0.24 \\
        & 2002 & 0.27 & 0.14 & 0.22 & 0.33 & 0.10 & 0.29 \\
        & 2003 & 0.17 & 0.13 & 0.08 & 0.09 & 0.08 & 0.09 \\
        & 2004 & 0.10 & 0.09 & 0.07 & 0.09 & 0.06 & 0.10 \\
        & 2005 & 0.12 & 0.11 & 0.15 & 0.17 & 0.09 & 0.15 \\
        & 2006 & 0.20 & 0.09 & 0.21 & 0.24 & 0.07 & 0.24 \\
        & 2007 & 0.32 & 0.21 & 0.08 & 0.16 & 0.21 & 0.10 \\
      & \textbf{Avg.} & \textbf{0.18} & \textbf{0.12} & \textbf{0.13} &
                       \textbf{0.17} & \textbf{0.10} & \textbf{0.16} \\
    \midrule
    \multirow{9}{*}{\textbf{\dypopm}}
        & 2001 & 0.29 & 0.15 & 0.16 & 0.28 & 0.13 & 0.24 \\
        & 2002 & 0.28 & 0.16 & 0.22 & 0.32 & 0.10 & 0.29 \\
        & 2003 & 0.31 & 0.17 & 0.08 & 0.16 & 0.13 & 0.09 \\
        & 2004 & 0.22 & 0.15 & 0.07 & 0.15 & 0.09 & 0.10 \\
        & 2005 & 0.22 & 0.16 & 0.15 & 0.19 & 0.10 & 0.15 \\
        & 2006 & 0.22 & 0.15 & 0.21 & 0.24 & 0.14 & 0.24 \\
        & 2007 & 0.30 & 0.16 & 0.08 & 0.14 & 0.09 & 0.10 \\
      & \textbf{Avg.} & \textbf{0.26} & \textbf{0.16} & \textbf{0.13} &
                       \textbf{0.21} & \textbf{0.12} & \textbf{0.16} \\
    \bottomrule
  \end{tabular}
\end{table}

Detailed in \autoref{tab:annual_performance}.


\begin{flushleft}
    \textbf{References Supplementary}
\end{flushleft}

{[}S1{]} Virtanen, P. et al. SciPy 1.0: Fundamental algorithms for
scientific computing in Python. \emph{Nat. Methods} \textbf{17},
261--272 (2020). https://doi.org/10.1038/s41592-019-0686-2

{[}S2{]} Petzold, L. R. Automatic selection of methods for solving stiff
and nonstiff systems of ordinary differential equations. \emph{SIAM J.
Sci. Stat. Comput.} 4, 136--148 (1983). https://doi.org/10.1137/0904010

\end{document}